\batchmode
\documentclass[twocolumn,english]{IEEEtran}
\usepackage[T1]{fontenc}
\usepackage{color}
\usepackage{babel}
\usepackage{amsmath}
\usepackage{amssymb}
\usepackage{graphicx}
\usepackage[unicode=true,
 bookmarks=true,bookmarksnumbered=true,bookmarksopen=true,bookmarksopenlevel=1,
 breaklinks=false,pdfborder={0 0 0},backref=false,colorlinks=false]
 {hyperref}
\hypersetup{pdftitle={Your Title},
 pdfauthor={Your Name},
 pdfpagelayout=OneColumn, pdfnewwindow=true, pdfstartview=XYZ, plainpages=false}

\makeatletter
 \let\oldforeign@language\foreign@language
 \DeclareRobustCommand{\foreign@language}[1]{%
   \lowercase{\oldforeign@language{#1}}}

\ifCLASSOPTIONcompsoc
\usepackage[caption=false,font=normalsize,labelfont=sf,textfont=sf]{subfig}
\else
\usepackage[caption=false,font=footnotesize]{subfig}
\fi

\makeatother

\begin{document}

\title{\textcolor{black}{Correlations of Interference and Link Successes
in Heterogeneous Cellular Networks}}

\author{Min~Sheng,~\IEEEmembership{Member,~IEEE,} Juan~Wen,~Jiandong~Li,~\IEEEmembership{Senior~Member,~IEEE,}
and~Ben~Liang,~\IEEEmembership{Senior~Member,~IEEE}%
\thanks{Min~Sheng, Juan Wen (Correspondence Author), and Jiandong Li are
with the State Key Laboratory of Integrated Service Networks, Institute
of Information Science, Xidian University, Xi'an, Shaanxi, 710071,
China. (e-mail: \{msheng, jdli\}@mail.xidian.edu.cn, juanwen66@gmail.com,
Ben Liang is with the Department of Electrical and Computer Engineering,
University of Toronto, Ontario, Canada. (e-mail: liang@comm.utoronto.ca).
Juan Wen is a visiting student at the University of Toronto supported
by the China Scholarship Council.%
}%
\thanks{This work was supported in part by National Natural Science Foundation
of China under Grant 61231008, 61172079, 61201141, 61301176, and 91338114,
by the National High Technology Research and Development Program of
China (863 Program) under Grant 2014AA01A701, by the 111 Project under
Grant B08038.%
}}

\markboth{}{Your Name \MakeLowercase{\emph{et al.}}: Your Title}
\maketitle
\begin{abstract}
\textcolor{black}{In heterogeneous cellular networks (HCNs), the interference
received at a user is correlated over time slots since it comes from
the same set of randomly located BSs. This results in the correlations
of link successes, thus affecting network performance.}\textcolor{red}{{}
}\textcolor{black}{Under the assumptions of a $K$-tier Poisson network,
strongest-candidate based BS association, and independent Rayleigh
fading, we first quantify the correlation coefficients of interference.
We observe that the interference correlation is independent of the
number of tiers, BS density, SIR threshold, and transmit power. Then,
we study the correlations of link successes in terms of the joint
success probability over multiple time slots. We show that the joint
success probability is decided by the success probability in a single
time slot and a diversity polynomial, which represents the temporal
interference correlation. Moreover, the parameters of HCNs have an
important influence on the joint success probability by affecting
the success probability in a single time slot. Particularly, we obtain
the condition under which the joint success probability increases
with the BS density and transmit power. We further show that the conditional
success probability given prior successes only depends on the path
loss exponent and the number of time slots.}\end{abstract}
\begin{IEEEkeywords}
heterogeneous cellular network, interference correlation, stochastic
geometry, joint success probability.
\end{IEEEkeywords}

\section{Introduction}

\IEEEPARstart{T}{o}\textcolor{red}{{} }\textcolor{black}{improve
the capacity of networks, heterogeneous cellular networks (HCNs) deploy
different kinds of irregular heterogeneous infrastructure elements,
such as micro, pico, and femtocells overlaid with traditional cellular
networks. In a traditional cellular network, without considering the
mobility, the interference received by a user is independent over
different time slots since the BSs are centrally planned and their
locations are given. However, in HCNs, due to the irregular deployment
of non-traditional BSs, the locations of these BSs are random. This
introduces correlation in the locations of BSs over time. As a result,
the interference in HCNs is temporally correlated even with independent
fading since the interference received by a user comes from the set
of randomly located and temporally correlated BSs. In this paper,
we focus on the interference correlation caused only by the random
locations of BSs.}

\textcolor{black}{The correlation in interference results in the correlation
in the success events of a link. For example, if a transmission succeeds
in the current time slot, there is a higher probability for successful
transmissions in the subsequent time slots. Such correlation affects
the performance of retransmission schemes and packet routing, thus
significantly impacting operations of networks. Therefore, it is important
to quantify the temporal correlations of interference and link successes. }

\textcolor{black}{However, in order to facilitate analytical tractability,
most prior literature assumes either complete correlation or no dependence,
which only captures the extremes. There exists prior works on interference
correlation, but they focus only on Poisson networks where the interferers
follow a single Poisson Point Process (PPP), instead of multi-tier
HCNs. In HCNs, each tier is distinguished by its BS density, transmit
power, and SIR threshold. It is clear that increasing the BS density
or transmit power will increase the interference power received by
the users. However, the following questions remain unanswered in HCNs:
do the number of tiers and the corresponding BS density, transmit
power, and SIR threshold affect the temporal correlations of interference
and link successes? If they do, what are their effects? These questions
are essential for the optimal design of HCNs and are the main focus
of this paper.}

We investigate the correlations of interference and link successes
in HCNs caused by the random BS locations. Assuming that the BSs in
HCNs are modeled as $K$ tiers of independently distributed PPPs and
the channels follow independent Rayleigh fading, we use the correlation
coefficient and joint success probability to quantify the correlations
of interference and link successes in HCNs, respectively. Futher,
we reveal the influences of some important system parameters on the
correlation coefficient and joint success probability. Based on the
result of joint success probability, the conditional success probability
in the $n^{th}$ time slot given successes in the prior $n-1$ time
slots is derived as another metric to quantify the correlations of
link successes.

\subsection{Related Work}

One of the most effective ways to improve wireless network capacity
is to increase the BS density by deploying low power BSs\cite{5G}.
Due to the random locations of such BSs, it is common to model them
as multi-tier PPPs, rather than traditional hexagonal grids, to analyze
the performance of HCNs \textcolor{black}{\cite{HCN-K-Tier,FlexibleCellAssociation,LoadAwareModeling,HCN-SINR,StructuredSpectrumAllocation,OffloadingHCNs,CapacityDownlinkMHCN}}.
Under the assumptions of independent Rayleigh fading and SINR threshold\textcolor{black}{{}
greater than one}, the authors in \cite{HCN-K-Tier} investigated
the performance of $K$-tier downlink HCNs in terms of the instantaneous
coverage probability and average rate, by calculating the complementary
cumulative distribution function of SINR. They observed that neither
the BS density nor the number of tiers changes the probability of
coverage or outage when all the tiers have the same SINR threshold.
Considering bias, the authors in \cite{FlexibleCellAssociation} derived
the outage probability and average rate in HCNs with full queues,
and the results are accurate at any SINR threshold. Using the maximum
biased received power association, the authors in \cite{OffloadingHCNs}
obtained that the optimum percentage of traffic to maximize SINR coverage
is different as that to maximize the rate coverage. Based on the idea
of conditionally thinning of the interference field, \cite{LoadAwareModeling}
relaxed the fully loaded assumption and computed the outage probability
in HCNs. However, all these prior works only obtained the success
probability or outage probability in a single time slot. The correlations
of interference and link successes in multiple time slots are not
considered.

Recently, researchers have started to pay attention to the interference
correlation caused by randomly located nodes since it severely affects
the performance of wireless networks. According to different configurations
for the receiver, the lines of recent literature can be divided into
three categories: the correlations between different time slots \cite{InterfCorreThreeSources,InterferenceCorreLetter,IntefCorrDiversityPolynomials,CorrelationMobileRandomNet},
the correlations between different receive antennas \cite{InterferenceCorrMRC},
and the correlations between different receivers \cite{IntefCorrMeanDelayReduce}.
The interference correlation was first investigated in ALOHA ad hoc
network whose nodes are distributed as a PPP \cite{InterferenceCorreLetter}.
It was shown that even with independent Rayleigh fading, there exist
correlations of interference and link successes since the interferers
come from the same random set. The authors used correlation coefficient
and joint success probability to quantify the correlations of interference
and link successes, respectively. However, the joint success probability
is not explicitly calculated. In \cite{IntefCorrDiversityPolynomials},
the expression of joint success probability in $n$ time slots was
obtained based on the diversity polynomial, which represents the temporal
interference correlation. The interference correlation caused by three
major sources, node locations, traffic and channel, was investigated
in \cite{InterfCorreThreeSources}. In \cite{InterfCorreDiversityLoss},
the impact of interference correlation on multi-antenna communication
was analyzed, and it was found that the probability of successful
reception over single-input multiple-output links is significantly
reduced by interference correlation. In that work, the closed-form
expression of the joint success probability is a special case of the
main result in \cite{IntefCorrDiversityPolynomials}. In \cite{CorrelationMobileRandomNet},
the temporal interference correlation in mobile Poisson networks was
quantified in terms of the correlation coefficient and conditional
outage probability. The results showed that smart routing, retransmission,
and multiple access control schemes are needed to avoid bursts of
transmission failures. However, all these analyses were conducted
on Poisson networks, where the interferers follow a PPP. Furthermore,
the distance between a transmitter and its receiver is assumed as
fixed. Thus, no cell association is considered in the prior work.

\subsection{Contributions}

In this paper, we focus on the problem whether and how the parameters
of HCNs, such as the number of tiers and the corresponding BS density,
transmit power, and SIR threshold affect the correlations of interference
and link successes. The main contributions of this paper are as follows:
\begin{itemize}
\item Using the tools of stochastic geometry, we derive the expressions
of the correlation coefficient, the joint success probability of $n$
transmissions, and the conditional success probability given $n-1$
prior successes under the assumptions of $K$ tiers of independent
PPP distributed BSs, Rayleigh fading channels, and SIR threshold\textcolor{black}{{}
greater than one}. Further, we obtain upper and lower bounds of the
joint success probability.
\item We find that the number of tiers and the corresponding BS density,
transmit power, and SIR threshold do not change the interference correlation
since they uniformly scales the interference.
\item The joint success probability is determined by two parts: the success
probability in a single time slot and the diversity polynomial which
represents the temporal interference correlation. Even if the change
of parameters in HCNs does not change the temporal interference correlation,
it affects the joint success probability since the success probability
in a single time slot is changed. More importantly, we provide the
condition under which the joint success probability is enhanced with
the increase of BS density and transmit power. Further, when all the
tiers have the same SIR threshold, the joint success probability remains
the same with the change of BS density and transmit power.
\item The conditional success probability in one time slot given that successes
occurred in the previous $n-1$ time slots is only decided by the
number of time slot and the path loss exponent.\\

\end{itemize}
The rest of the paper is organized as follows. Section II describes
the system model used in this paper. The correlation of interference
is derived in Section III. Section IV investigates the joint success
probability and conditional success probability in HCNs to quantify
the correlations of link successes and reveals the effect of SIR threshold,
BS density and transmit power on the joint success probability. Section
V presents the numerical results. Finally, the conclusion is given
in Section VI.

\begin{figure}[t]
\centering\includegraphics[scale=0.5]{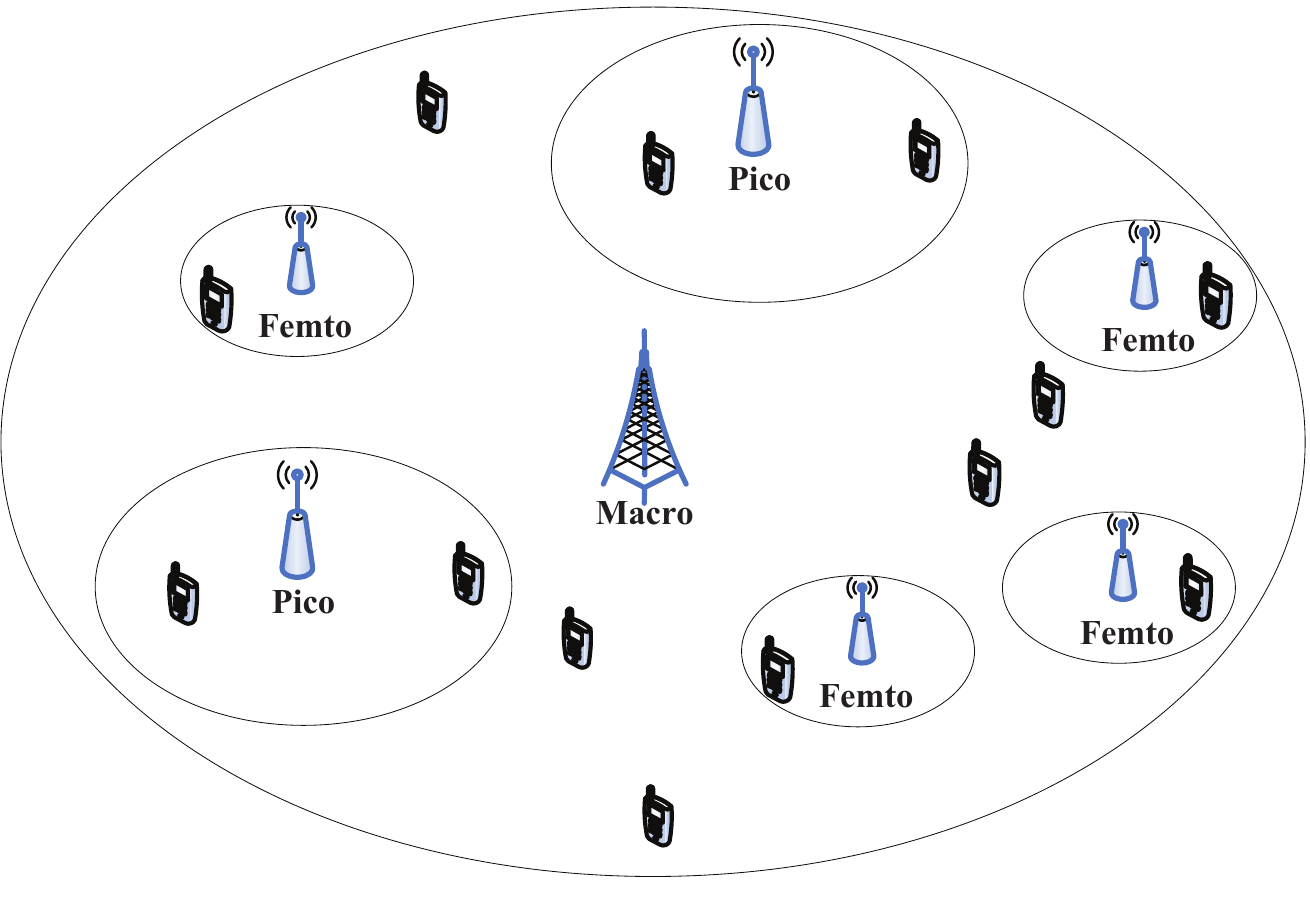}\caption{Illustration of heterogeneous cellular network composed of macro,
pico and femtocell BSs and users.}
\end{figure}

\section{System Model}

Fig. 1 shows a HCN consisting of three tiers of BSs (macro, pico and
femtocell BSs) and mobile users. The scenario of multiple macro-cells
is considered in our analysis although we only show a single macro-cell
for the sake of simplicity. In this paper, we model downlink HCNs
as $K$-tier BSs that are distinguished by their BS density $\lambda_{k}$,
transmit power $P_{k}$, and SIR threshold $\beta_{k}$. The locations
of BSs in each tier independently follow homogeneous PPP $\phi_{k}\,\left(1\leq k\leq K\right)$.
The fading between BSs and any mobile user is assumed as temporally
and spatially independent Rayleigh fading and the power fading coeffi{}cient
between BS $x$ and the typical mobile user located at origin in time
slot $t$ is denoted by $h_{x}(t)\sim exp(1).$ The standard singular
path loss function is given as $g(x)=\frac{1}{\Vert x\Vert^{\alpha}},$
where $\alpha>2$ is the path loss exponent in $d$-dimensional plane.
The analysis is conducted on a single frequency band. We assume that
all BSs transmit packets continuously in all time slots at constant
power. With the above notations, the interference of the typical user
located at the origin and associated with BS $x_{i}$ in time slot
$t$ is expressed as
\begin{equation}
I_{t}\left(x_{i}\right)=\sum_{l=1}^{K}\sum_{x\in\phi_{l},x\neq x_{i}}P_{l}h_{x}\left(t\right)g\left(x\right).
\end{equation}
\textcolor{black}{Since we assume noise is negligible, the received
SIR of the user is accordingly given by}
\begin{equation}
SIR_{t}\left(x_{i}\right)=\frac{P_{i}h_{x_{i}}\left(t\right)g\left(x_{i}\right)}{\sum_{l=1}^{K}\sum_{x\in\phi_{l},x\neq x_{i}}P_{l}h_{x}\left(t\right)g\left(x\right)}.\label{eq:SIR}
\end{equation}
We consider an open access strategy \cite{HCN-K-Tier}, in which a
mobile user can successfully associate with a BS in the $k^{th}$
tier only if its SIR with respect to that BS is greater than the corresponding
threshold $\beta_{k}$. \textcolor{black}{Under the assumption that
$\beta_{k}>1,\:\forall k$, at most one BS in the entire network can
provide SIR greater than the required threshold \cite[Lemma 1]{HCN-K-Tier}. }

\section{\textcolor{black}{The Correlations of Interference in HCNs}}

Due to the stationarity of PPP, all users have the same interference
distribution when the transmitters follow PPPs. Therefore, we may
conduct analysis on a typical user located at the origin without loss
of generality. However, interference is not independent across the
plane or time slots. This is because interference is caused by the
same random point processes \cite{InterferenceCorreLetter}. In this
section, we will investigate the spatio-temporal correlation of interference
in HCNs and derive the correlation coefficient.

It has been previously observed that, when the standard singular path
loss function $g(x)=\frac{1}{\Vert x\Vert^{\alpha}},\:\alpha>2$,
is used, the average interference and its higher moments are infinite
\cite{Interference-LargeNet}. In this case, the correlation coefficient
is undefined. Instead, we first use a bounded path loss function $g_{\varepsilon}\left(x\right)=\frac{1}{\Vert x\Vert^{\alpha}+\varepsilon},\:\varepsilon\in\left(0,\infty\right),\:\alpha>2$,
to calculate the correlation coefficient and then consider the limiting
case when $\varepsilon\downarrow0$. The following Theorem gives the
spatial and temporal correlation coefficient of interference in HCNs.

\textbf{Theorem 1.} \emph{The correlation coefficient of interference
in HCNs where the BSs follow $K$-tier PPPs is
\begin{equation}
\rho\left(I_{t_{1}}(u),I_{t_{2}}(v)\right)=\frac{\int_{\mathbb{\mathbb{R}}^{d}}g(x)g(x-\Vert u-v\Vert)dx}{\mathbb{E}\left[h^{2}\right]\int_{\mathbb{\mathbb{R}}^{d}}g^{2}(x)dx}.
\end{equation}
Further, the temporal correlation coefficient of interference }(\emph{$\Vert u-v\Vert=0$})\emph{
is expressed as
\begin{equation}
\rho_{t}=\frac{1}{\mathbb{E}\left[h^{2}\right]}.
\end{equation}
The spatial correlation coefficient is given by
\begin{equation}
\underset{\varepsilon\downarrow0}{\lim}\rho\left(I_{t}(u),I_{t}(v)\right)=0,\; u\neq v.
\end{equation}
}
\begin{IEEEproof}
See Appendix A.
\end{IEEEproof}
From Theorem 1, we can see that the correlation coefficient of interference
is independent of the number of tiers, and the corresponding BS density
and transmit power. \textcolor{black}{To explain this unintuitive
result, we consider a HCN without fading or mobility. In this case,
the interference power received by a typical user remains the same
for all time slots and the corresponding correlation coefficient is
one. Different realizations of BSs, such as different number of tiers
and the corresponding BS density and transmit power, will change the
interference power at all time slots, without affecting the interference
correlation. This is because they uniformly scale the interference.}
We also obtain that the correlation coefficient in HCNs is the same
as that of ALOHA ad hoc networks (where the transmitters follow a
PPP) with the ALOHA selection probability $p=1$. This is because
the summation of several independent PPPs is still a PPP.

For $u=v$ and $t_{1}\neq t_{2}$, we obtain the temporal correlation
coefficient which is only dependent on fading channels. If the channels
are subject to independent Rayleigh fading with parameter 1, the temporal
correlation coefficient is equal to 0.5. When $u\neq v$ and $t_{1}=t_{2}$,
we obtain the spatial correlation coefficient for the standard singular
path loss function when $\varepsilon\downarrow0$. It should be noted
that the spatial correlation coefficient being 0 is an artifact. The
reason is as follows. When $g\left(x\right)=\frac{1}{\Vert x\Vert^{\alpha}}$,
the interference created by user $u$ is mainly decided by the transmitters
in a disc $B\left(u,r\right)$ centered at $u$ with a small radius
$r>0$. In PPPs, the transmitters locations of different user in $B\left(u,r\right)$
and $B\left(v,r\right)$ are independent with each other for a small
$r$. Therefore, the correlation coefficient goes to zero.

\section{\textcolor{black}{The Correlations of Link Successes in HCNs}}

In this section, joint success probabiltiy and conditional success
probability are obtained to quantify the correlations of link successes
in HCNs. The joint success probability is defined as the probability
that a typical user accesses to a BS with SIR above its corresponding
threshold in $n$ successive time slots. Under the assumption of SIR
threshold $\beta_{k}>1\: for\, all\: k$, a user can connect to at
most one BS among all BSs in the entire network at any given time
slot. Furthermore, since no mobility is considered in this paper,
a typical user always connects to the same BS in the $n$ time slots.
\textcolor{black}{According to the law of total probability, the joint
success probability in HCNs is the summation of the probability that
the user connects to every BS in HCNs. In other words, we have }
\begin{equation}
p^{(n)}=\mathbb{E}_{\phi}\left[\sum_{i=1}^{K}\sum_{x_{i}\in\phi_{i}}p_{x_{i}}^{(n)}\right],
\end{equation}
where $p_{x_{i}}^{(n)}$ denotes the probability that a typical user
accesses to a given BS $x_{i}$ in $n$ successive time slots.

To derive $p^{(n)}$, we first calculate $p_{x_{i}}^{(n)}$.\textcolor{black}{{}
We also obtain some properties of the joint success probability. The
conditional success probability is further derived based on the joint
success probability.}

\subsection{Joint Success Probability of HCNs}

\textbf{\textcolor{black}{Lemma 1. }}\emph{The probability that a
typical user located at the origin accesses to the given BS $x_{i}$
in $n$ successive time slots is expressed as
\begin{gather}
p_{x_{i}}^{(n)}=P^{x_{i}}\left(A\left(t_{1}\right),A\left(t_{2}\right),\cdots,A\left(t_{n}\right)\right)\nonumber \\
=\exp\left(-c_{d}\frac{\pi\delta}{\sin\pi\delta}\left(\frac{\beta_{i}}{P_{i}}\right)^{\delta}\sum_{l=1}^{K}\lambda_{l}P_{l}^{\delta}D_{n}\left(\delta\right)\cdot\Vert x_{i}\Vert^{d}\right),
\end{gather}
where $A\left(t_{n}\right)$ represents the success event in $t_{n}$,
$c_{d}$ is the volume of the $d$-dimensional unit ball, $\delta=\frac{d}{\alpha}$,
and $D_{n}\left(\delta\right)$ denotes diversity polynomial \cite{InterfCorreDiversityLoss}
which is the multivariable polynomial given by $D_{n}\left(\delta\right)=\frac{\Gamma\left(n+\delta\right)}{\Gamma\left(n\right)\Gamma\left(1+\delta\right)}$
. }
\begin{IEEEproof}
See Appendix B.
\end{IEEEproof}
Lemma 1 gives a general expression for \emph{$d$-}dimensional joint
success probability of a user associated with the given BS \emph{$x_{i}$
}in $n$ time slots. When $d=2$, the result is expressed as $\exp\left(-\pi\frac{\pi\delta}{\sin\pi\delta}\left(\frac{\beta_{i}}{P_{i}}\right)^{\delta}\sum_{l=1}^{K}\lambda_{l}P_{l}^{\delta}D_{n}\left(\delta\right)\cdot\Vert x_{i}\Vert^{2}\right)$.
When $d=2$ and $K=1$, the result in Lemma 1 can be simplified to
the special case of homogeneous ad hoc networks where the interferers
follow a PPP. In this case, it is easy to see that we obtain the same
result in \cite{InterfCorreDiversityLoss} directly by setting $d=2,$
and $K=1$.

Under the assumption of $\beta_{i}>1,\:\forall i$, a user can access
to at most one BS in the whole network. Therefore, the joint success
probability in HCNs can be calculated by the sum of the probabilities
that the user connects to each BS in $n$ time slots. This is because
all the events are mutually exclusive and cannot happen at the same
time. Note that the sum of probabilities over PPPs can be converted
to a simple integral using Campbell-Mecke Theorem \cite{StoGeo1-Theory}.
The expression of joint success probability in $n$ time slots is
derived in the following Theorem.

\textbf{Theorem 2.} \emph{Given $\beta_{i}>1\:\forall i$, the joint
success probability for a randomly located typical user in $n$ successive
time slots is}
\begin{eqnarray}
p^{(n)} & \overset{\triangle}{=} & P\left(A\left(t_{1}\right),A\left(t_{2}\right),\cdots,A\left(t_{n}\right)\right)\nonumber \\
 & = & \frac{\sum_{i=1}^{K}\lambda_{i}P_{i}^{\delta}\beta_{i}^{-\delta}}{\frac{\pi\delta}{\sin\pi\delta}\cdot D_{n}\left(\delta\right)\cdot\sum_{l=1}^{K}\lambda_{l}P_{l}^{\delta}}.\label{eq:JointSuccProb}
\end{eqnarray}

\begin{IEEEproof}
Due to the spatial stationarity of PPP, all the users have the same
statistics of received signal \cite{StoGeo1-Theory}. Therefore, the
analysis can be conducted on a typical user located at the origin.
The joint success probability is:
\begin{flalign*}
p^{(n)} & \overset{\triangle}{=}P\left(A\left(t_{1}\right),A\left(t_{2}\right),\cdots,A\left(t_{n}\right)\right)\\
\overset{\left(a\right)}{=} & \mathbb{E}_{\phi}\left[\sum_{i=1}^{K}\sum_{x_{i}\in\phi_{i}}P^{x_{i}}\left(A\left(t_{1}\right),A\left(t_{2}\right),\cdots,A\left(t_{n}\right)\right)\right]\\
\overset{\left(b\right)}{=} & \sum_{i=1}^{K}\mathbb{E}_{\phi}\left[\sum_{x_{i}\in\phi_{i}}P^{x_{i}}\left(A\left(t_{1}\right),A\left(t_{2}\right),\cdots,A\left(t_{n}\right)\right)\right]\\
\overset{\left(c\right)}{=} & \sum_{i=1}^{K}\int_{\mathbb{R}^{d}}\lambda_{i}\\
\cdot & \exp\left(-c_{d}\frac{\pi\delta}{\sin\pi\delta}\left(\frac{\beta_{i}}{P_{i}}\right)^{\delta}\sum_{l=1}^{K}\lambda_{l}P_{l}^{\delta}D_{n}\left(\delta\right)\Vert x_{i}\Vert^{d}\right)dx_{i}\\
= & \sum_{i=1}^{K}c_{d}d\lambda_{i}\int_{0}^{\infty}\\
 & \exp\left(-c_{d}\frac{\pi\delta}{\sin\pi\delta}\left(\frac{\beta_{i}}{P_{i}}\right)^{\delta}\sum_{l=1}^{K}\lambda_{l}P_{l}^{\delta}D_{n}\left(\delta\right)r^{d}\right)r^{d-1}dr\\
= & \frac{\sum_{i=1}^{K}\lambda_{i}P_{i}^{\delta}\beta_{i}^{-\delta}}{\frac{\pi\delta}{\sin\pi\delta}\cdot D_{n}\left(\delta\right)\cdot\sum_{l=1}^{K}\lambda_{l}P_{l}^{\delta}},
\end{flalign*}
where $\left(a\right)$ comes from the assumption that $\beta_{i}>1\:\forall i$,
$\left(b\right)$ follows from the linearity of the expectation, $\left(c\right)$
comes from the Campbell-Mecke Theorem.
\end{IEEEproof}
Although the result is not a closed-form expression, it is still a
simple expression including the diversity polynomial $D_{n}\left(\delta\right)$
which can be calculated by using the gamma function. According to
the properties of $D_{n}\left(\delta\right)$\cite{InterfCorreDiversityLoss},
the upper and lower bound of joint success probability in $n$ successive
time slots is obtained in Corollary 1.

\textbf{Corollary 1.} \emph{The upper bound of the joint success probability
is $\frac{1}{\frac{\pi\delta}{\sin\pi\delta}n^{\delta}}\cdot\frac{\sum_{i=1}^{K}\lambda_{i}P_{i}^{\delta}\beta_{i}^{-\delta}}{\sum_{l=1}^{K}\lambda_{l}P_{l}^{\delta}}$
and the lower bound is }$\frac{\Gamma\left(1+\delta\right)}{\frac{\pi\delta}{\sin\pi\delta}n^{\delta}}\cdot\frac{\sum_{i=1}^{K}\lambda_{i}P_{i}^{\delta}\beta_{i}^{-\delta}}{\sum_{l=1}^{K}\lambda_{l}P_{l}^{\delta}}$.\emph{ }
\begin{IEEEproof}
Note that $\alpha>d$, $\delta=\frac{d}{\alpha}$, so $0<\delta<1$.
For all $\delta\in\left(0,1\right)$, $\frac{\Gamma\left(n+\delta\right)}{\Gamma\left(n\right)}\text{<}n^{\delta}$.
From \cite{InterfCorreDiversityLoss}, we find that
\begin{equation}
n^{\delta}<D_{n}\left(\delta\right)<\frac{n^{\delta}}{\Gamma\left(1+\delta\right)}.\label{eq:Bound_D}
\end{equation}
Substituting (\ref{eq:Bound_D}) into Theorem 1, we obtain the bounds
of joint success probability.
\end{IEEEproof}
Next, we study the relationship between the joint success probability
in $n$ time slots $p^{(n)}$ and the success probability in a single
time slot $p^{(1)}$.\emph{ }Since $D_{1}\left(\delta\right)=1$,
$p^{(1)}$ is expressed as
\begin{equation}
p^{(1)}=\frac{1}{\frac{\pi\delta}{\sin\pi\delta}}\cdot\frac{\sum_{i=1}^{K}\lambda_{i}P_{i}^{\delta}\beta_{i}^{-\delta}}{\sum_{l=1}^{K}\lambda_{l}P_{l}^{\delta}},\label{eq:SuccProP1}
\end{equation}
which coincides with the result in \cite{HCN-K-Tier}\emph{.} Comparing
(\ref{eq:JointSuccProb}) and (\ref{eq:SuccProP1}), the joint success
probability can be expressed as $p^{(n)}=\frac{1}{D_{n}(\delta)}\cdot p^{(1)}.$
Therefore, $p^{(n)}$ is determined by both $p^{(1)}$ and the diversity
polynomial $D_{n}(\delta)$ representing temporal correlation.\emph{ }

From the expression of $D_{n}(\delta)$, we see that the success event
are fully correlated when $\delta\downarrow0$ $\left(\alpha\uparrow\infty\right)$
and independent when $\delta\uparrow1$ $\left(\alpha\downarrow2\right)$.
If $\delta\downarrow0$ $\left(\alpha\uparrow\infty\right)$, $D_{n}\left(\delta\right)\downarrow1$
for all $n$, so $p^{(1)}=p^{(2)}=\cdots=p^{(n)}$. In this case,
the success events are fully correlated. The reason is that the interference
is dominated by some near-by interferers. If they cause an outage
at the current time slot, it is likely to do so at the next time slots.
On the other hand, if $\delta\uparrow1$ $\left(\alpha\downarrow2\right)$,
$D_{n}\left(\delta\right)\uparrow n$, the success events become independent.
This is because the interference is influenced by many faraway interferers.
The fading states between the interferers and the user in different
time slots are independent, leading to independence in the corresponding
interference. It is worth noting that at the same time, we have $p^{(n)}\downarrow0$
since $\sin\left(\frac{2\pi}{\alpha}\right)\downarrow0$ when $\alpha\downarrow2$.

According to (\ref{eq:JointSuccProb}), the joint success probability
in $n$ time slots is affected by BS density, SIR threshold, and transmit
power. The following two corollaries tell us how the system parameters
influence the joint success probability.

\textbf{Corollary 2. }\emph{Given the path loss exponent $\alpha$
and SIR threshold $\beta_{i}$, when the given SIR thresholds of various
tiers are not all the same, the increase in BS density or transmit
power improves the joint success probability under the condition that
$\beta_{m}^{-\delta}>\frac{\sum_{i=1,i\neq m}^{K}\lambda_{i}P_{i}^{\delta}\beta_{i}^{-\delta}}{\sum_{i=1,i\neq m}^{K}\lambda_{i}P_{i}^{\delta}}$
. Otherwise, the increase in BS density or transmit power reduces
the joint success probability.}
\begin{IEEEproof}
To prove the joint success probability $p^{(n)}$ increases with the
BS density $\lambda_{m}$, we denote the constants $C_{0}=\frac{1}{\frac{\pi\delta}{\sin\pi\delta}D_{n}\left(\delta\right)}$,
$C_{1}=\sum_{i=1,i\neq m}^{K}\lambda_{i}P_{i}^{\delta}\beta_{i}^{-\delta}$,
$C_{2}=\lambda_{m}\beta_{m}^{-\delta}$, $D_{1}=\sum_{i=1,i\neq m}^{K}\lambda_{i}P_{i}^{\delta}$,
$D_{2}=\lambda_{m}$. The joint success probability is expressed as
\begin{equation}
p^{(n)}=C_{0}\cdot\frac{C_{1}+C_{2}P_{m}^{\delta}}{D_{1}+D_{2}P_{m}^{\delta}}.
\end{equation}
Taking the derivative with respect to $P_{m}$, we obtain
\begin{equation}
p^{\prime(n)}=C_{0}\cdot\frac{\delta\cdot P_{m}^{\delta-1}\left(C_{2}D_{1}-C_{1}D_{2}\right)}{\left[D_{1}+D_{2}\lambda_{m}\right]^{2}}.
\end{equation}
Since $C_{0}>0$, $\delta>0$, and $P_{m}^{\delta-1}>0$, we have
$p^{\prime(n)}>0$ when $C_{2}D_{1}>C_{1}D_{2}$. Thus, when $\beta_{m}^{-\delta}>\frac{\sum_{i=1,i\neq m}^{K}\lambda_{i}P_{i}^{\delta}\beta_{i}^{-\delta}}{\sum_{i=1,i\neq m}^{K}\lambda_{i}P_{i}^{\delta}}$,
the joint success probability $p^{(n)}$ increases with the transmit
power $P_{m}$ of the $m^{th}$ tier. On the other side, when $C_{2}D_{1}<C_{1}D_{2}$,
we have $p^{\prime(n)}<0$. Therefore, the increase in transmit power
$P_{m}$ of the $m^{th}$ tier will decrease the joint success probability
under the condition that $\beta_{m}^{-\delta}<\frac{\sum_{i=1,i\neq m}^{K}\lambda_{i}P_{i}^{\delta}\beta_{i}^{-\delta}}{\sum_{i=1,i\neq m}^{K}\lambda_{i}P_{i}^{\delta}}$.

We can also show that the joint success probability increases with
increasing BS density under the same condition, in a similar approach
as the above, by denoting $C_{2}=P_{m}^{\delta}\beta_{m}^{-\delta}$
and $D_{2}=P_{m}^{\delta}$. This completes the proof.
\end{IEEEproof}
\textcolor{black}{An intuitive explanation for Corollary 2 is the
following. We know that increasing BS density or transmit power will
increase the received power and the interference power at the same
time. When the condition in Corollary 2 holds, the increase in the
received power is greater than the increase in the interference power
since the users have a high probability to access to the BSs in the
$m^{th}$ tier. This is because the smaller the value of the SIR threshold
of tier $m$ is, the higher probability the users will connect to
the BSs in that tier, and the easier the above condition will hold
at the same time. Therefore, the increase in transmit power and BS
density of tier $m$ will result in the improvement of the joint success
probability. Conversely, if the condition in Corollary 2 does not
hold, the joint success probability is reduced by increasing the BS
density or transmit power due to the inter-cell interference.}\textcolor{red}{{}
}We should note that Corollary 2 is obtained under the assumption
that the given SIR thresholds of various tiers are not all the same.
The next corollary tells us how the system parameters affect the joint
success probability when all the tiers have the same SIR threshold.

\textbf{Corollary 3.} \emph{When all the tiers have the same SIR threshold
($\beta_{i}=\beta$), the joint success probability $p^{(n)}$ is
expressed as
\begin{equation}
p^{(n)}=\frac{1}{\frac{\pi\delta}{\sin\pi\delta}D_{n}\left(\delta\right)\beta^{\delta}}.\label{eq:JointSuccProb_SameBeta}
\end{equation}
which is only decided by the same SIR threshold $\beta$ and diversity
polynomial $D_{n}\left(\delta\right)$ .}
\begin{IEEEproof}
We obtain (\ref{eq:JointSuccProb_SameBeta}) by substituting $\beta_{i}=\beta,\:\forall i$
into Theorem 1.
\end{IEEEproof}
\textcolor{black}{When all tiers have the same SIR threshold, the
users will choose the BSs depending only on the received SIR from
that BS and the common SIR threshold. It means that the users cannot
differentiate the BSs belonging to different tiers. Thus, the change
to the number of tiers and their relative transmit power and BS density
results in a change in interference power and signal power with the
same factor. The corresponding effects are also canceled. Therefore,
the joint success probability is only decided by SIR threshold $\beta$
and diversity polynomial $D_{n}\left(\delta\right)$ which represents
temporal correlation.} When there is a strong temporal correlation
($\alpha\uparrow\infty$), the joint success probability is high.
Otherwise, the joint success probability is low. Especially, when
$\alpha\downarrow2$, the joint success probability approximates to
0.

Next, the conditional success probability given $n-1$ successes is
obtained as another metric to quantify the correlations of link successes.

\subsection{\textcolor{black}{The Conditional Success Probability}}

\textcolor{black}{In this section, we investigate the conditional
success probability which is defined as the probability that the $n$
attempt succeeds when the first $n-1$ ones did. Note that it is another
metric to quantify the correlations of link successes. }

\textbf{Corollary 4.} \emph{The conditional success probability given
n successes in HCNs is expressed as
\begin{flalign}
 & P\left(A_{t_{n}}|A_{t_{1}},A_{t_{2}},\cdots,A_{t_{n-1}}\right)\nonumber \\
= & \frac{P\left(A_{t_{1}},A_{t_{2}},\cdots,A_{t_{n}}\right)}{P\left(A_{t_{1}},A_{t_{2}},\cdots,A_{t_{n-1}}\right)}\label{eq:ConditionalSuccProb}\\
= & \frac{D_{n-1}\left(\delta\right)}{D_{n}\left(\delta\right)},\nonumber
\end{flalign}
which is decided only by the number of time slots n and path loss
exponent $\alpha$ ($\delta$ is determined by $\alpha$ and $\delta=\frac{d}{\alpha}$).}
\begin{IEEEproof}
We obtain this result directly by substituting (\ref{eq:JointSuccProb})
into (\ref{eq:ConditionalSuccProb}).
\end{IEEEproof}
From Corollary 4, we see that the conditional success probability
is independent of BS density, transmit power, and SIR threshold and
it only depends on the path loss exponent $\alpha$ and the number
of time slots $n$. For $n\uparrow\infty$, we get that $P\left(A_{t_{n}}|A_{t_{1}},A_{t_{2}},\cdots,A_{t_{n-1}}\right)\uparrow1$.
It shows that the $n$ attempt will succeed with probability 1 when
the first $n-1$ ones succeed.

\subsection{Comparison with the Special Cases in the Existing Work}

In this section, we will derive the joint success probability of some
special cases that have been studied in existing works, by setting
specific system parameters in the above analysis. Further, the relationship
between them and our result is obtained.

\subsubsection{SIMO System}

When $K=1$, the system in our paper can be simplified to a SIMO system
where the interferers come from a stationary Poisson point process
and the receiver under consideration is equipped with $n\geq1$ antennas
\cite{InterfCorreDiversityLoss}. In this case, the SIR in different
time slots in our paper should be considered as the SIR at different
antennas in \cite{InterfCorreDiversityLoss}. It is worth noting that
in \cite{InterfCorreDiversityLoss}, the authors assume that the desired
transmitter is added at a distance $r$ from the receiver, which is
different from our system. Therefore, there is no cell association
in \cite{InterfCorreDiversityLoss}. We obtain the probability that
the SIR at all antennas exceeds SIR threshold $\beta$ \cite[Theorem 1]{InterfCorreDiversityLoss}
by setting $d=2$, $c_{d}=\pi$, $P=1$, $\parallel x_{i}\parallel=r$
in Lemma 1 \textcolor{black}{since we do not consider cell association
in Lemma 1}. The corresponding probability is expressed as $\exp\left(-\frac{\pi^{2}\delta}{\sin\left(\pi\delta\right)}\cdot\lambda\cdot r^{2}\cdot\beta^{\delta}\cdot D_{n}\left(\delta\right)\right)$.
Based on the property of the gamma function $\Gamma\left(1-\delta\right)\Gamma\left(1+\delta\right)=\frac{\pi\delta}{\sin\left(\pi\delta\right)}$,
the above probability is the same as the result in \cite[Theorem 1]{InterfCorreDiversityLoss}.
Note that the diversity polynomial $D_{n}\left(\delta\right)$ quantifies
the spatial diversity in \cite{InterfCorreDiversityLoss} instead
of the temporal diversity in our paper.

\subsubsection{Ad hoc Networks}

When $K=1$, our system model can also be seemed as an ad hoc network
where the transmitters follow a stationary PPP. In particular, this
special case is the same as the ALOHA ad hoc network investigated
in \cite{InterferenceCorreLetter} with transmission probability $p=1$.
In \cite{InterferenceCorreLetter}, the distance between the transmitter
and its receiver is assumed as a constant, which is different to our
model. We can obtain the joint success probability in \cite{InterferenceCorreLetter}
by setting $d=2$, $c_{d}=\pi$, $P=1$, $\parallel x_{i}\parallel=r$
and $n=2$ in Lemma 1. This is because the cell association is not
cosindered in Lemma 1. It is worth noting that the authors in \cite{InterferenceCorreLetter}
did not derive the closed-form expression.

\subsubsection{HCNs without considering temporal correlations}

If we only consider a single time slot ($n=1$), there is no interference
correlation. Correspondingly, the diversity polynomial $D_{n}\left(\delta\right)$
equals to one. Therefore, the success probability of HCNs in one time
slot is expressed as $p^{(1)}=\frac{1}{\frac{\pi\delta}{\sin\pi\delta}}\cdot\frac{\sum_{i=1}^{K}\lambda_{i}P_{i}^{\delta}\beta_{i}^{-\delta}}{\sum_{l=1}^{K}\lambda_{l}P_{l}^{\delta}}$
which coincides with the result in \cite[Corollary 1]{HCN-K-Tier}
with no noise\emph{. }Further, if $K=1$ or all the tiers have the
same SIR threshold $\beta_{i}=\beta,\:\forall i$, the success probability
in a single time slot is simplified to $p^{(1)}=\frac{1}{\frac{\pi\delta}{\sin\pi\delta}}\cdot\beta^{-\delta}$,
which is only decided by the SIR threshold and the path loss exponent.

\subsection{Special Case of Interest }

To avoid cross-tier interference, one option is to allocate separated
spectrum to the BSs in different tiers. In this special case, the
interference received by a typical user comes from the BSs in the
same tier and is expressed as $I_{t}\left(x_{i}\right)=\sum_{x\in\phi_{i},x\neq x_{i}}P_{i}h_{x}\left(t\right)g\left(x\right)$.
According to the result in Lemma 1, the probability that a typical
user connects to the given BS $x_{i}$ in $n$ successive time slots
is given by
\begin{equation}
p_{x_{i}}^{(n)}=\exp\left(-c_{d}\frac{\pi\delta}{\sin\pi\delta}\lambda_{i}\beta_{i}^{\delta}D_{n}\left(\delta\right)\cdot\Vert x_{i}\Vert^{d}\right).
\end{equation}
It is worth noting that if we consider the two-dimensional space,
the above expression is the same as the result of Theorem 1. in \cite{IntefCorrDiversityPolynomials}.
This is because the interference only comes from a PPP, and fixed
distance between transmitter and receiver is considered.

Next, based on the assumption of $\beta_{i}>1,\:\forall i$, and the
Campbell-Mecke Theorem, we obtain the joint success probability that
the typical user accesses to the BSs in tier $i$ as
\begin{equation}
p_{i}^{(n)}=\frac{\beta_{i}^{-\delta}}{\frac{\pi\delta}{\sin\pi\delta}D_{n}\left(\delta\right)}.
\end{equation}
The joint success probability of a typical tier is only dependent
on the SIR threshold of the corresponding tier and the path loss exponent.

\section{Numerical Results}

In this section, using Matlab and Monte Carlo methods, we first validate
the correlation coefficient and joint success probability in Theorem
1 and Theorem 2, respectively. Then, the effect of SIR threshold,
BS density, and transmit power on the joint success probability is
discussed.

In the simulations, the locations of BSs in each tier follow a PPP.
The typical user is assumed at the origin. The fading channels are
generated as independent Rayleigh random variables. The user chooses
the strongest BS in terms of the received SIR and the SIR \textcolor{black}{is
evaluated by} (\ref{eq:SIR}). In each Monte Carlo trial, the channel
gains and the locations of BSs are generated independently.

\begin{figure}[tbh]
\centering\includegraphics[width=8.8cm,height=5.7cm]{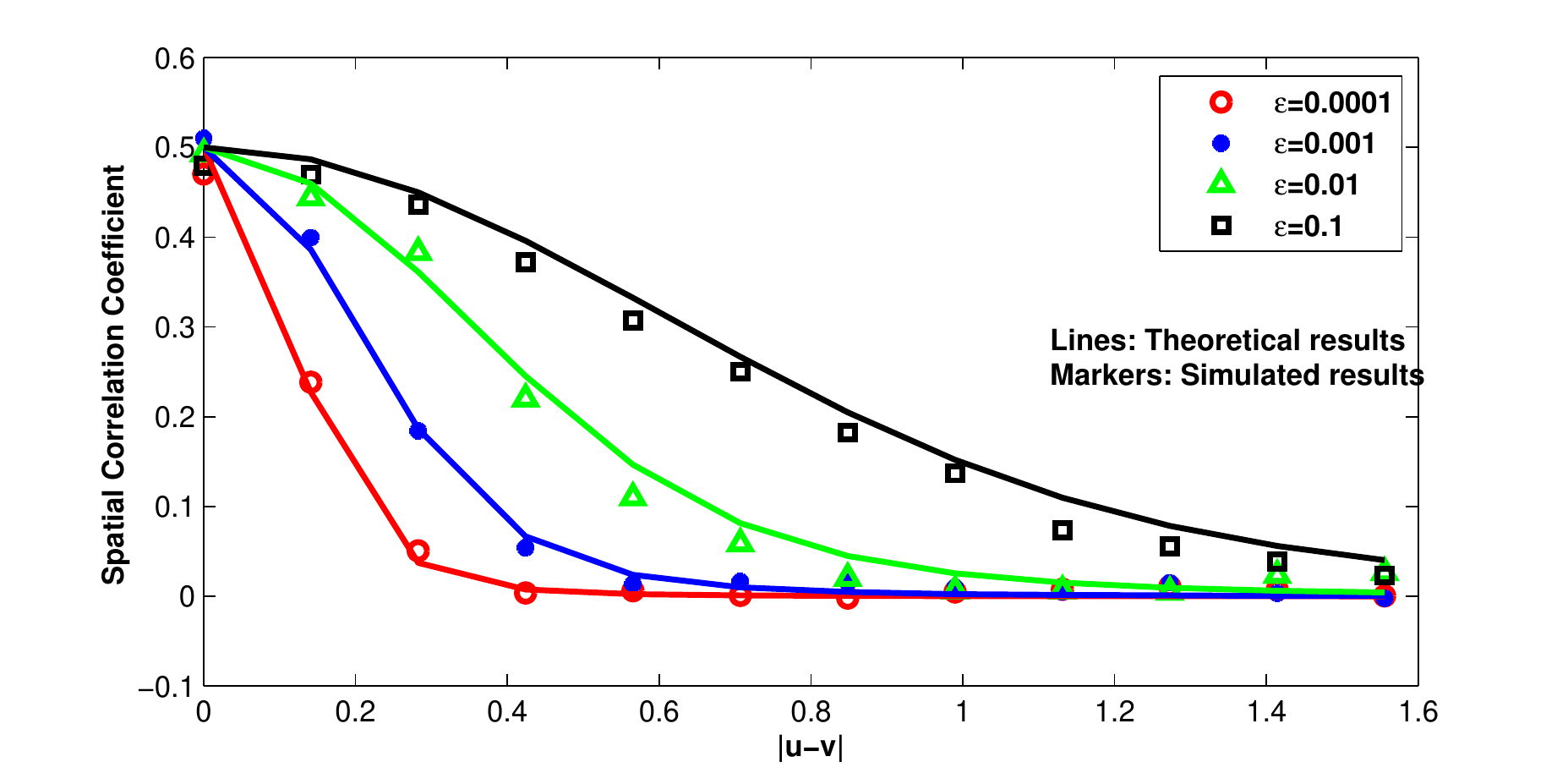}\caption{Correlation coefficient of interference versus $\Vert u-v\Vert$}
\label{Fig_CorrelationCoefficient}
\end{figure}

Fig. \ref{Fig_CorrelationCoefficient} shows the correlation coefficient
of interference varying with $\Vert u-v\Vert$ for $\alpha=4$, $\mathbb{E}\left[h^{2}\right]=2$,
and $g_{\varepsilon}\left(x\right)=\frac{1}{\Vert x\Vert^{\alpha}+\varepsilon},\:\varepsilon$
taking small positive values. We observe that the correlation coefficient
of interference reaches the maximum value when $\Vert u-v\Vert=0$.
This gives the temporal correlation coefficient at the same location.
Note that it is only dependent on $\mathbb{E}\left[h^{2}\right]$
since $\rho_{t}=\frac{1}{\mathbb{E}\left[h^{2}\right]}$. The correlation
coefficient decreases with the increase of $\Vert u-v\Vert$, which
coincides with our intuition. The farther the distance is, the smaller
the correlation coefficient is. When $\varepsilon\downarrow0$, the
correlation coefficient goes to zero. This is because the interference
is dominated by the transmitters located near to the receiver and
the interferers for different receivers are independent in PPPs.

Fig. 3 shows the joint success probability of two-tier HCNs represented
by simulated results, analytical results, the corresponding upper
and lower bound, and the independent case. From Fig. 3, we first notice
that an analysis that ignores the interference correlation (i.e.,
the independent case) can substantially under estimate the joint success
probability. We further observe that the simulated results coincide
with the analytical results very well even for $\beta_{2}=-4dB$.
The upper and lower bounds are also close to the analytical results.
Further, we observe that the joint success probability decreases with
the increase of SIR threshold. This is because it is harder for the
users to obtain satisfactory SIR when the SIR threshold increases.

\begin{figure}[tbh]
\centering\includegraphics[width=8.8cm,height=5.7cm]{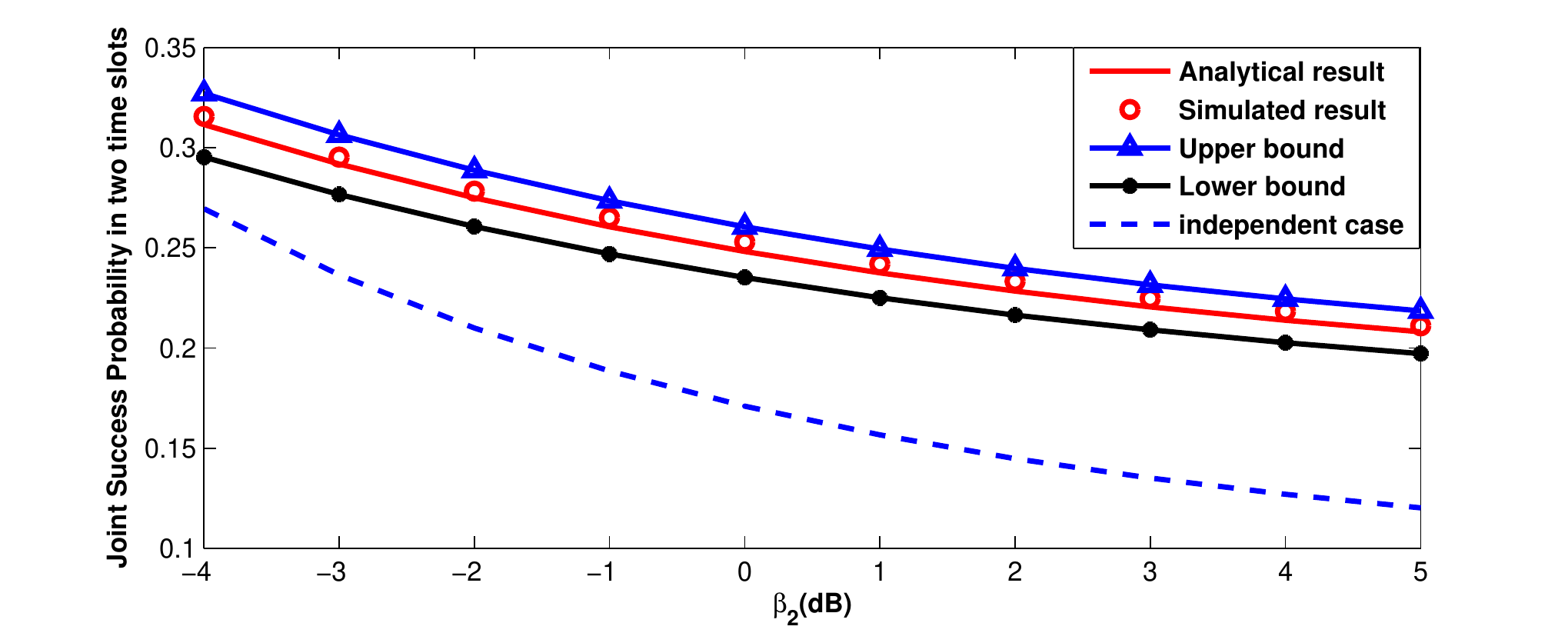}\label{Fig.JointSuccProb_beta}

\caption{Joint success probability in a two-tier HCNs ($n=2$, $K=2$, $P_{1}=10P_{2}$,
$\lambda_{2}=2\lambda_{1}$, $\beta_{1}=0dB$, $\alpha=3$).}
\end{figure}

Fig. 4 and Fig. 5 show the influence of BS density and transmit power
of the second tier on the joint success probability. When $K=2$,
the condition in Corollary 2 is simplified to $\beta_{2}<\beta_{1}$.
When $\beta_{2}=-4dB$ and $\beta_{2}=-2dB$, the above condition
holds ($\beta_{2}<\beta_{1}$). Hence, the joint success probability
is improved by increasing the BS density and transmit power of tier
two. The reason is that the users prefer to access to the BSs in the
second tier when the condition holds. Thus, increasing the BS density
or transmit power leads to a higher increase in the received power
than that in the interference. When the condition in Corollary 2 does
not hold, the joint success probability decreases with increasing
the BS density or transmit power, since the resulting increase in
the received power is less than that in the interference. When $\beta_{2}=1dB$
($\beta_{1}=\beta_{2}$), the joint success probability does not vary
with the change of BS density or transmit power. This is because the
change in the above system parameters results in the same change in
the received power and the interference and thus the effect can be
canceled.

\begin{figure}[tbh]
\centering\includegraphics[width=8.8cm,height=5.7cm]{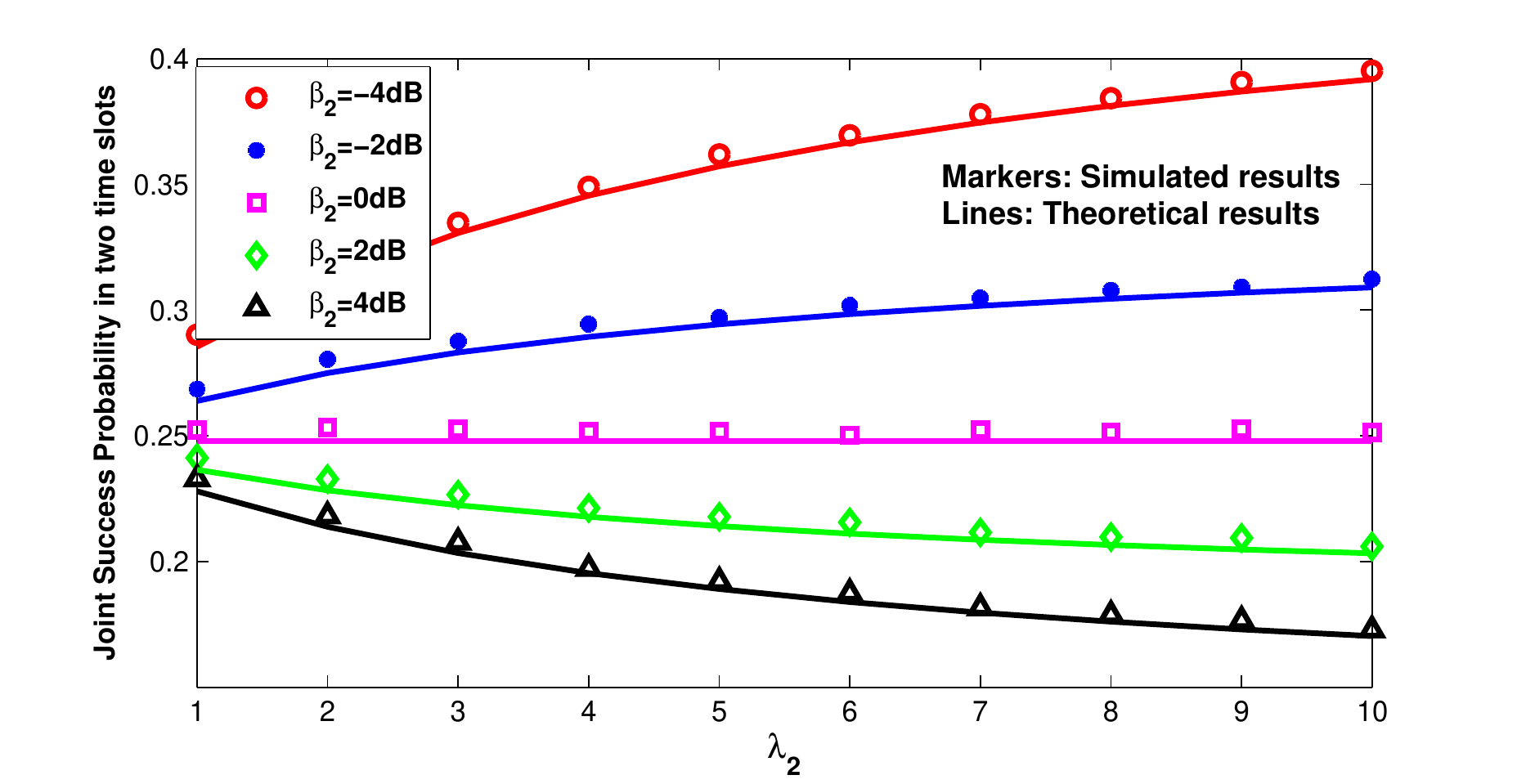}\label{Fig.JointSuccPro_lamda}\caption{Joint success probability in a two-tier HCNs ($n=2,$ $K=2$, $P_{1}=10P_{2}$,
$\lambda_{1}=1$, $\beta_{1}=0dB$, $\alpha=3$)}
\end{figure}

\begin{figure}[tbh]
\centering\includegraphics[width=8.8cm,height=5.7cm]{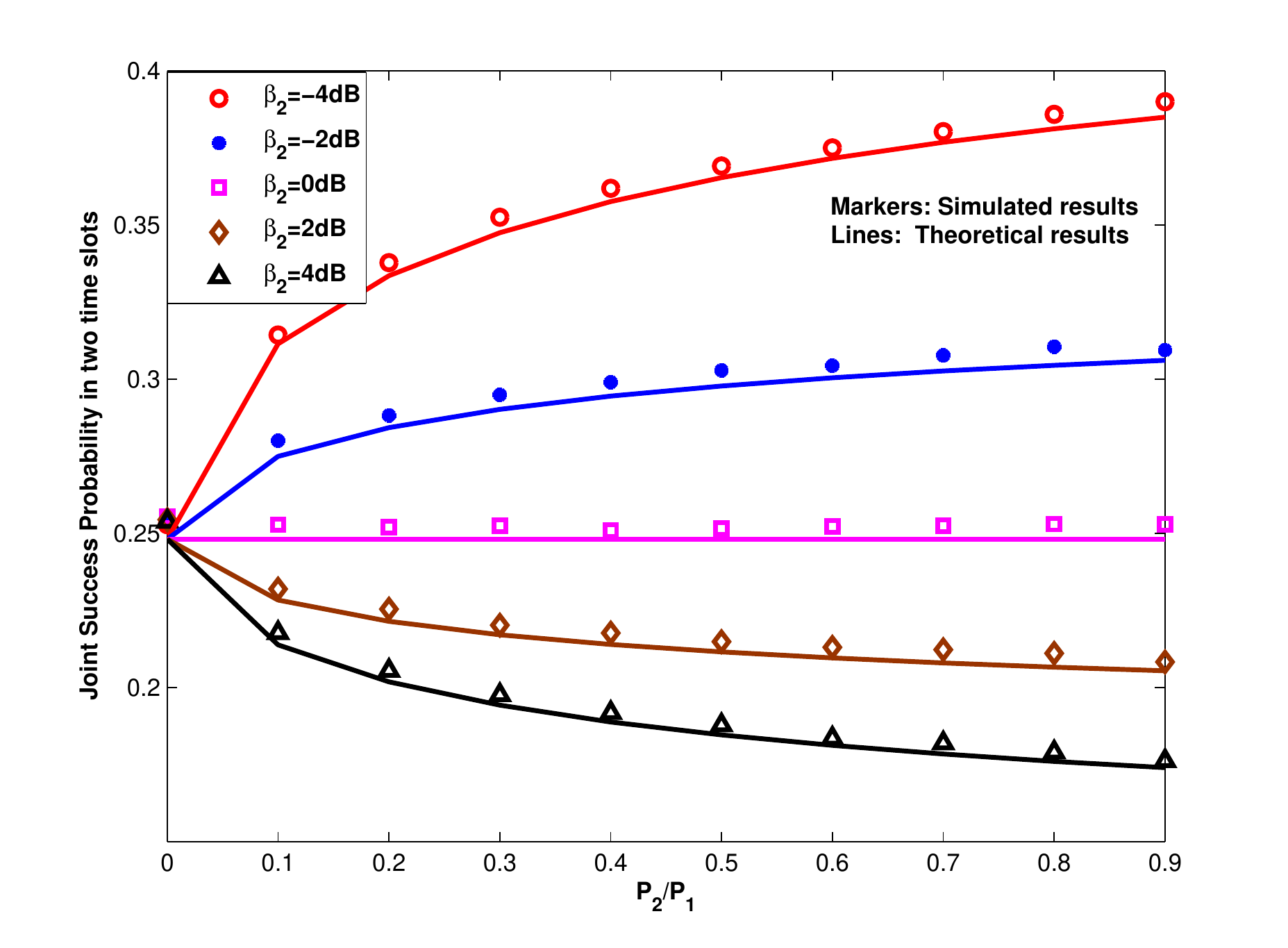}\label{Fig.JointSuccPro_Pt}

\caption{Joint success probability in a two-tier HCNs ($n=2,$ $K=2$, $P_{1}=100$,
$\lambda_{2}=2\lambda_{1}$, $\beta_{1}=0dB$, $\alpha=3$)}
\end{figure}

In Fig. 6, we present the conditional success probability in a two-tier
HCNs varying with the number of time slots. From this figure, we see
that the conditional success probability is dramatically enhanced
in the first and second slots. Further, the conditional success probability
approximates to 1 when $n\rightarrow\infty$. At the same time slot,
the greater the path loss exponent is, the higher the conditional
success probability is, which coincides with our intuition. The interference
is mainly determined by some nearest interferers when there is a large
path loss exponent. Therefore, the successes in the previous time
slots will lead to success in the current time slot with a high probability.

\begin{figure}[tbh]
\centering\includegraphics[width=8.8cm,height=5.7cm]{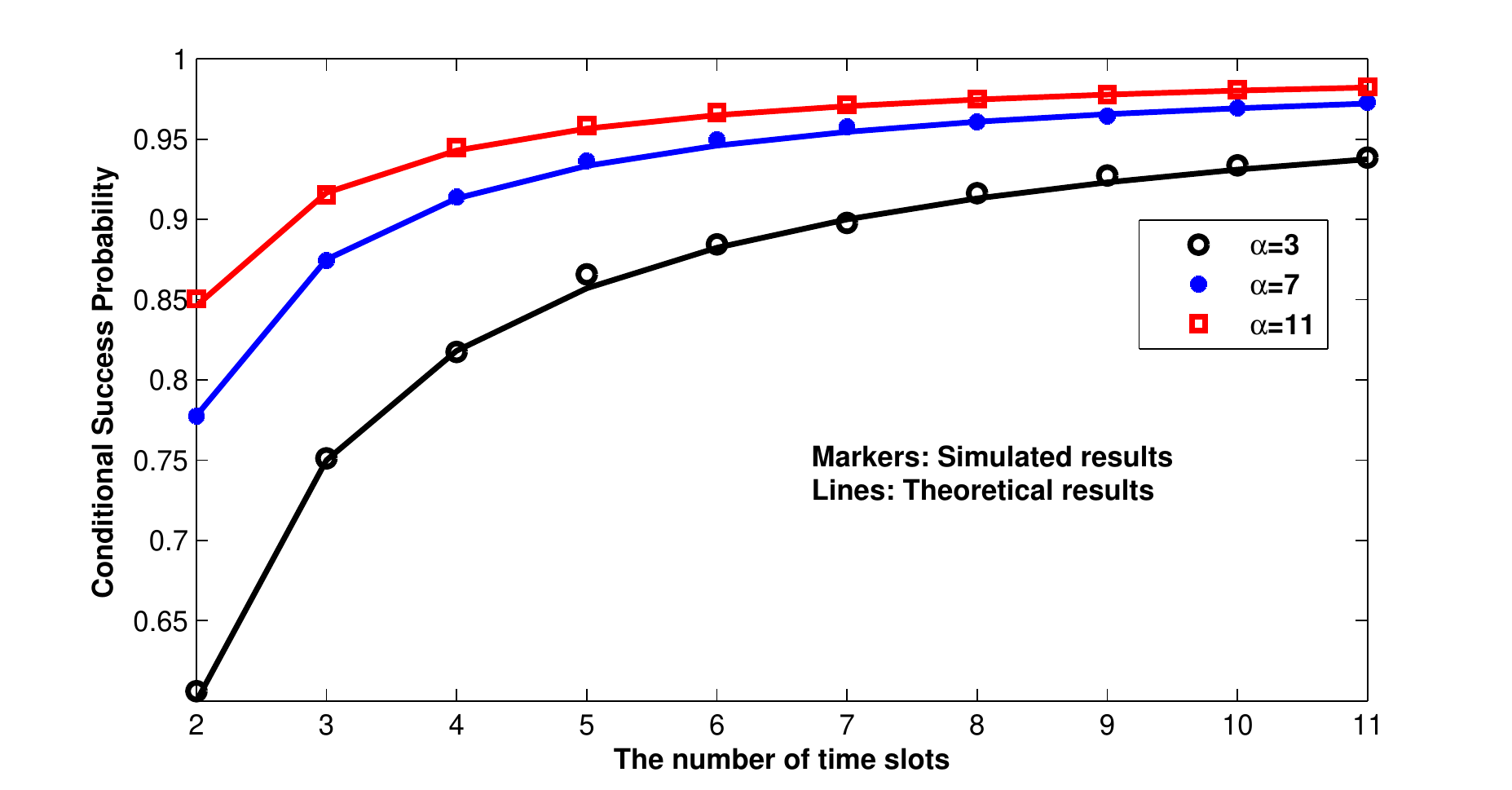}\label{Fig.Conditional Probability}\caption{Conditional success probability in a two-tier HCNs ($n=2,$ $K=2$,
$P_{1}=10P_{2}$, $\lambda_{2}=2\lambda_{1}$, $\beta_{1}=0dB$)}
\end{figure}

\section{Conclusions}

In this paper, we investigated the correlations of interference and
link successes which are quantified by the correlation coefficient
and joint success probability in HCNs. Based on the $K$-tier Poisson
network model, Rayleigh fading, and SIR threshold greater than one,
we derived the expressions of the above metrics. From our analytical
results, we revealed how the system parameters affect the interference
correlation and the joint success probability. We showed that the
interference correlation coefficient is independent of the number
of tiers, BS density, and transmit power. Further, we observed that
the temporal correlation coefficient is dependent only on channel
fading. However, although the parameters of HCNs do not change the
correlations of interference, they have an important influence on
the correlations of link outages. When the SIR thresholds are not
all the same, joint success probability is enhanced with the increase
in BS density or transmit power under a certain condition. When all
tiers have the same SIR threshold, the joint success probability is
only decided by the diversity polynomial and the SIR threshold. Finally,
we obtained the conditional success probability after $n-1$ successes,
which is another metric to quantify the correlations of link successes.
We observed that the conditional success probability is only dependent
on the number of time slots and path loss exponent.

\appendices{}

\section{Appendix: Proofs}

\subsection{Proof of Theorem 1}
\begin{IEEEproof}
According to our system model, the interference of a randomly selected
user located at $u$ at time slot $t$ is expressed as $I_{t}\left(u\right)=\sum_{l=1}^{K}\sum_{x\in\phi_{l},x\neq x_{i}}P_{l}h_{x}\left(t\right)g\left(x-u\right).$
Since all users have the same interference distribution, we can conduct
the analysis on a typical user located at the origin. The average
interference is given by
\begin{flalign}
\mathbb{E}\left[I_{t}\left(u\right)\right] & =\mathbb{E}\left[I_{t}\left(o\right)\right]\nonumber \\
 & =\mathbb{E}\left[\sum_{j=1}^{K}\sum_{x\in\phi_{j},x\neq x_{u}}P_{j}h_{x}\left(t\right)g\left(x\right)\right]\nonumber \\
 & \overset{\left(a\right)}{=}\sum_{j=1}^{K}\mathbb{E}_{\phi_{j}}\left[\sum_{x\in\phi_{j},x\neq x_{u}}\mathbb{E}_{h}\left[P_{j}h_{x}\left(t\right)g\left(x\right)\right]\right]\nonumber \\
 & \overset{\left(b\right)}{=}\sum_{j=1}^{K}\lambda_{j}\int_{\mathbb{\mathbb{R}}^{d}}\mathbb{E}[h]P_{j}g(x)dx\nonumber \\
 & =\sum_{j=1}^{K}\lambda_{j}P_{j}\int_{\mathbb{\mathbb{R}}^{d}}g(x)dx,\label{eq:Expectation}
\end{flalign}
where $\left(a\right)$ comes from the linearity of the expectation,
$\left(b\right)$ follows from Campbell-Mecke Theorem.

The mean product of $I_{t_{1}}(u)$ and $I_{t_{2}}(v)$ at different
time slots $t_{1}$and $t_{2}$ is given by
\begin{flalign}
 & \mathbb{E}[I_{t_{1}}(u),I_{t_{2}}(v)]\nonumber \\
 & =\mathbb{E}[\sum_{j=1}^{K}\sum_{\underset{x\neq x_{u}}{x\in\phi_{j}}}P_{j}h_{x}(t_{1})g(x-u)\sum_{q=1}^{K}\sum_{\underset{x\neq y_{v}}{y\in\phi_{q}}}P_{q}h_{y}(t_{2})g(y-v)]\nonumber \\
 & =\mathbb{E}[\sum_{j=1}^{K}\sum_{\underset{x\neq x_{u}}{x\in\phi_{j}}}P_{j}h_{x}(t_{1})g(x-u)\cdot P_{j}h_{x}(t_{2})g(x-v)\nonumber \\
 & +\sum_{j=1}^{K}\sum_{\underset{x\neq x_{u},x\neq y}{x\in\phi_{j}}}\sum_{q=1}^{K}\sum_{\underset{x\neq y_{v},x\neq y}{y\in\phi_{q}}}P_{j}h_{x}(t_{1})g(x-u)\cdot P_{q}h_{y}(t_{2})g(y-v)]\nonumber \\
 & \overset{\left(a\right)}{=}\sum_{j=1}^{K}P_{j}^{2}(\mathbb{E}[h])^{2}\lambda_{j}\int_{\mathbb{\mathbb{R}}^{d}}g(x-u)g(x-v)dx\nonumber \\
 & +\sum_{j=1}^{K}\sum_{q=1}^{K}P_{j}P_{q}(\mathbb{E}[h])^{2}\lambda_{j}\lambda_{q}\int_{\mathbb{\mathbb{R}}^{d}}g(x-u)dx\int_{\mathbb{\mathbb{R}}^{d}}g(y-v)dy\nonumber \\
 & =\sum_{j=1}^{K}P_{j}^{2}\lambda_{j}\int_{\mathbb{\mathbb{R}}^{d}}g(x-u)g(x-v)dx\nonumber \\
 & +\sum_{j=1}^{K}\sum_{q=1}^{K}P_{j}P_{q}\lambda_{j}\lambda_{q}(\int_{\mathbb{\mathbb{R}}^{d}}g(x)dx)^{2},\label{eq:Mean Product}
\end{flalign}
where $\left(a\right)$follows by the second order product density
of PPPs and Campbell's theorem.

The second moment of the interference is expressed as
\begin{flalign}
\mathbb{E}\left[I_{t}^{2}(u)\right] & =\mathbb{E}\left[I_{t}^{2}(o)\right]\nonumber \\
 & =\mathbb{E}\left[\left(\sum_{j=1}^{K}\sum_{x\in\phi_{j},x\neq x_{u}}P_{j}h_{x}(t)g(x)\right)^{2}\right]\nonumber \\
 & =\mathbb{E}\left[\sum_{j=1}^{K}\sum_{x\in\phi_{j},x\neq x_{u}}P_{j}^{2}h_{x}^{2}(t)g^{2}(x)\right]\nonumber \\
 & +\mathbb{E}\left[\sum_{j=1}^{K}\sum_{\underset{x\neq x_{u},x\neq y}{x\in\phi_{j}}}\sum_{q=1}^{K}\sum_{\underset{x\neq y_{v},x\neq y}{y\in\phi_{q}}}P_{j}h_{x}(t)g(x)\cdot P_{q}h_{y}(t)g(y)\right]\nonumber \\
 & =\sum_{j=1}^{K}P_{j}^{2}\mathbb{E}\left[h^{2}\right]\lambda_{j}\int_{\mathbb{\mathbb{R}}^{d}}g^{2}(x)dx\nonumber \\
 & +\sum_{j=1}^{K}\sum_{q=1}^{K}P_{j}P_{q}(\mathbb{E}[h])^{2}\lambda_{j}\lambda_{q}(\int_{\mathbb{\mathbb{R}}^{d}}g(x)dx)^{2}.
\end{flalign}
The variance of the interference is given by
\begin{flalign}
VAR\left[I_{t}(u)\right] & =\mathbb{E}[I_{t}^{2}(u)]-\left(\mathbb{E}[I_{t}(u)]\right)^{2}\nonumber \\
 & =\sum_{j=1}^{K}P_{j}^{2}\mathbb{E}\left[h^{2}\right]\lambda_{j}\int_{\mathbb{\mathbb{R}}^{d}}g^{2}(x)dx.\label{eq:Variance}
\end{flalign}
The correlation coefficient of two random variables is expressed as
\begin{flalign}
\rho\left(X_{1},X_{2}\right) & =\frac{cov\left(X_{1},X_{2}\right)}{\sqrt{var\left(X_{1}\right)}\cdot\sqrt{var\left(X_{1}\right)}}\nonumber \\
 & =\frac{\mathbb{E}\left[X_{1},X_{2}\right]-\mathbb{E}\left[X_{1}\right]\mathbb{E}\left[X_{2}\right]}{\sqrt{var\left(X_{1}\right)}\cdot\sqrt{var\left(X_{1}\right)}}.\label{eq:OriCorCoefficient}
\end{flalign}
Substituting (\ref{eq:Mean Product}), (\ref{eq:Expectation}), and
(\ref{eq:Variance}) into (\ref{eq:OriCorCoefficient}), we obtain
the spatial-temporal correlation coefficient of the interference $I_{t_{1}}(u)$
and $I_{t_{2}}(v)$ that
\begin{flalign}
\rho\left(I_{t_{1}}(u),I_{t_{2}}(v)\right) & =\frac{\int_{\mathbb{\mathbb{R}}^{d}}g(x)g(x-\Vert u-v\Vert)dx}{\mathbb{E}\left[h^{2}\right]\int_{\mathbb{\mathbb{R}}^{d}}g^{2}(x)dx}.
\end{flalign}
The temporal correlation coefficient is obtained as $\rho_{t}=\frac{1}{\mathbb{E}\left[h^{2}\right]}$
by setting $\Vert u-v\Vert=0$. Note that the above derivation is
obtained when $g\left(x\right)$ is defined as a bounded path loss
function $g_{\varepsilon}\left(x\right)=\frac{1}{\Vert x\Vert^{\alpha}+\varepsilon},\:\varepsilon\in\left(0,\infty\right),\:\alpha>d$.
However, we obtain the correlation coefficient for the singular path
loss as $\varepsilon\downarrow0$. The spatial correlation coefficient
is given by $\underset{\varepsilon\downarrow0}{\lim}\rho\left(u,v\right)=0,\; u\neq v$
\cite{InterferenceCorreLetter}.
\end{IEEEproof}

\subsection{Proof of Lemma 1}
\begin{IEEEproof}
Recall that the fading is averaged out, a user connects to the strongest
BS in terms of the long-term averaged received power. Since no mobility
is considered in our paper, the user is associated with the same BS
in $n$ successive time slots. Given BS $x_{i}$, the joint success
probability of a typical user located at the origin is:
\begin{flalign*}
p_{x_{i}}^{(n)}= & P^{x_{i}}\left(A\left(t_{1}\right),A\left(t_{2}\right),\cdots,A\left(t_{n}\right)\right)\\
= & \mathbb{P}\left(SIR_{t_{1}}(x_{i})>\beta_{i},SIR_{t_{2}}(x_{i})>\beta_{i},\cdots,SIR_{t_{n}}(x_{i})>\beta_{i}\right)\\
= & \mathbb{P}\left(\frac{P_{i}h_{x_{i}}(t_{1})g(x_{i})}{I_{t_{1}}(x_{i})}>\beta_{i},\cdots,\frac{P_{i}h_{x_{i}}(t_{n})g(x_{i})}{I_{t_{n}}(x_{i})}>\beta_{i}\right)\\
= & \mathbb{P}\left(h_{x_{i}}(t_{1})>\frac{\beta_{i}I_{t_{1}}(x_{i})}{P_{i}g(x_{i})},\cdots,h_{x_{i}}(t_{n})>\frac{\beta_{i}I_{t_{n}}(x_{i})}{P_{i}g(x_{i})}\right)\\
\overset{\left(a\right)}{=} & \mathbb{E}_{I_{t}}\left[\exp\left(-\frac{\beta_{i}\left(I_{t_{1}}(x_{i})+I_{t_{2}}(x_{i})+\cdots I_{t_{n}}(x_{i})\right)}{P_{i}g(x_{i})}\right)\right]\\
\overset{\left(b\right)}{=} & \mathbb{E}_{I_{t}}\left[\exp\left(-\frac{\beta_{i}P_{l}g(x)}{P_{i}g(x_{i})}\sum_{l=1}^{K}\sum_{,\underset{x\neq x_{i}}{x\in\phi_{l}}}\left(h_{x}(t_{1})+\cdots+h_{x}\left(t_{n}\right)\right)\right)\right]\\
=\Pi_{l=1}^{K} & \mathbb{E}_{\phi_{l}}\left[\underset{\underset{x\neq x_{i}}{x\in\phi_{l}}}{\Pi}\mathbb{E}_{h}\left[\exp\left(-\frac{\beta_{i}P_{l}g(x)}{P_{i}g(x_{i})}\left(h_{x}(t_{1})+\cdots+h_{x}\left(t_{n}\right)\right)\right)\right]\right]\\
\overset{\left(c\right)}{=} & \Pi_{l=1}^{K}\mathbb{E}_{\phi_{l}}\left[\underset{\underset{x\neq x_{i}}{x\in\phi_{l}}}{\Pi}\left(\frac{1}{1+\frac{\beta_{i}P_{l}g(x)}{P_{i}g(x_{i})}}\right)^{n}\right]\\
\overset{\left(d\right)}{=} & \Pi_{l=1}^{K}\exp\left(-\lambda_{l}\int_{\mathbb{R}^{d}}\left[1-\left(\frac{1}{1+\frac{\beta_{i}P_{l}g(x)}{P_{i}g(x_{i})}}\right)^{n}\right]dx\right)\\
\overset{\left(e\right)}{=} & \exp\left(-c_{d}\frac{\pi\delta}{\sin\pi\delta}\left(\frac{\beta_{i}}{P_{i}}\right)^{\delta}\sum_{l=1}^{K}\lambda_{l}P_{l}^{\delta}D_{n}\left(\delta\right)\cdot\Vert x_{i}\Vert^{d}\right),
\end{flalign*}
where $\left(a\right)$ comes from the independence of $h_{x_{i}}(t_{1})$
, $h_{x_{i}}(t_{2})$,$\cdots$, $h_{x_{i}}(t_{n})$ , $\left(b\right)$
follows from the expression of interference $I_{t}\left(x_{i}\right)=\sum_{l=1}^{K}\sum_{x\in\phi_{l},x\neq x_{i}}P_{l}h_{x}\left(t\right)g\left(x\right)$,
$\left(c\right)$ follows by taking the average with respect to $h_{x}(t_{1})$
, $h_{x}(t_{2})$,$\cdots$, $h_{x}(t_{n})$, $\left(d\right)$ comes
from probability generating functional of PPP, and $\left(e\right)$
follows from the calculation of the integral \cite{InterfCorreDiversityLoss}.
\end{IEEEproof}
\bibliographystyle{IEEEtran}

\begin{thebibliography}{10}
\providecommand{\url}[1]{#1}
\csname url@samestyle\endcsname
\providecommand{\newblock}{\relax}
\providecommand{\bibinfo}[2]{#2}
\providecommand{\BIBentrySTDinterwordspacing}{\spaceskip=0pt\relax}
\providecommand{\BIBentryALTinterwordstretchfactor}{4}
\providecommand{\BIBentryALTinterwordspacing}{\spaceskip=\fontdimen2\font plus
\BIBentryALTinterwordstretchfactor\fontdimen3\font minus
  \fontdimen4\font\relax}
\providecommand{\BIBforeignlanguage}[2]{{%
\expandafter\ifx\csname l@#1\endcsname\relax
\typeout{** WARNING: IEEEtran.bst: No hyphenation pattern has been}%
\typeout{** loaded for the language `#1'. Using the pattern for}%
\typeout{** the default language instead.}%
\else
\language=\csname l@#1\endcsname
\fi
#2}}
\providecommand{\BIBdecl}{\relax}
\BIBdecl

\bibitem{5G}
K.~Mallinson, ``The 2020 vision for {LTE},''
  http://www.fiercewireless.com/europe/story/mallinson-2020-vision-lte/2012-06-20,
  Tech. Rep.

\bibitem{HCN-K-Tier}
H.S.Dhillon, R.K.Ganti, F.Baccelli, and J.G.Andrews, ``Modeling and analysis of
  k-tier downlink heterogeneous cellular networks,'' \emph{IEEE J. Sel. Areas
  Commun.}, vol.~30, no.~3, pp. 550--560, April 2012.

\bibitem{FlexibleCellAssociation}
H.-S. Jo, Y.~J. Sang, P.~Xia, and J.~Andrews, ``Heterogeneous cellular networks
  with flexible cell association: A comprehensive downlink sinr analysis,''
  \emph{IEEE Trans. Wireless Commun.}, vol.~11, no.~10, pp. 3484--3495, 2012.

\bibitem{LoadAwareModeling}
H.~Dhillon, R.~Ganti, and J.~Andrews, ``Load-aware modeling and analysis of
  heterogeneous cellular networks,'' \emph{IEEE Trans. Wireless Commun.},
  vol.~12, no.~4, pp. 1666--1677, 2013.

\bibitem{HCN-SINR}
S.~Mukherjee, ``Distribution of downlink sinr in heterogeneous cellular
  networks,'' \emph{IEEE J. Sel. Areas Commun.}, vol.~30, no.~3, pp. 575--585,
  April 2012.

\bibitem{StructuredSpectrumAllocation}
W.~Bao and B.~Liang, ``Structured spectrum allocation and user association in
  heterogeneous cellular networks,'' in \emph{Proc. IEEE INFOCOM}, 2014, pp.
  1069--1077.

\bibitem{OffloadingHCNs}
S.~Singh, H.~Dhillon, and J.~Andrews, ``Offloading in heterogeneous networks:
  Modeling, analysis, and design insights,'' \emph{IEEE Trans. Wireless
  Commun.}, vol.~12, no.~5, pp. 2484--2497, 2013.

\bibitem{CapacityDownlinkMHCN}
J.~Wen, M.~Sheng, X.~Wang, J.~Li, and H.~Sun, ``On the capacity of downlink
  multi-hop heterogeneous cellular networks,'' \emph{IEEE Trans. Wireless
  Commun.}, vol.~13, no.~8, pp. 4092--4103, Aug 2014.

\bibitem{InterfCorreThreeSources}
U.~Schilcher, C.~Bettstetter, and G.~Brandner, ``Temporal correlation of
  interference in wireless networks with rayleigh block fading,'' \emph{IEEE
  Trans. Mobile Comp.}, vol.~11, no.~12, pp. 2109--2120, Dec 2012.

\bibitem{InterferenceCorreLetter}
R.~Ganti and M.~Haenggi, ``Spatial and temporal correlation of the interference
  in aloha ad hoc networks,'' \emph{IEEE Commun. Lett.}, vol.~13, no.~9, pp.
  631--633, Sept 2009.

\bibitem{IntefCorrDiversityPolynomials}
M.~Haenggi and R.~Smarandache, ``Diversity polynomials for the analysis of
  temporal correlations in wireless networks,'' \emph{IEEE Trans. Wireless
  Commun.}, vol.~12, no.~11, pp. 5940--5951, November 2013.

\bibitem{CorrelationMobileRandomNet}
Z.~Gong and M.~Haenggi, ``Interference and outage in mobile random networks:
  Expectation, distribution, and correlation,'' \emph{IEEE Trans. Mobile
  Comp.}, vol.~13, no.~2, pp. 337--349, 2014.

\bibitem{InterferenceCorrMRC}
R.~Tanbourgi, H.~S. Dhillon, J.~G. Andrews, and F.~K. Jondral, ``Effect of
  spatial interference correlation on the performance of maximum ratio
  combining,'' \emph{IEEE Trans. Wireless Commun.}, July, 2013.

\bibitem{IntefCorrMeanDelayReduce}
Y.~Zhong, W.~Zhang, and M.~Haenggi, ``Managing interference correlation through
  random medium access,'' \emph{IEEE Trans. Wireless Commun.}, vol.~PP, no.~99,
  pp. 1--14, 2014.

\bibitem{InterfCorreDiversityLoss}
M.Haenggi, ``Diversity loss due to interference correlation,'' \emph{IEEE
  Commun. Lett.}, vol.~16, no.~10, pp. 1600--1603, October 2012.

\bibitem{Interference-LargeNet}
M.~Haenggi and R.~Ganti, \emph{Interference in large wireless netwoks}.\hskip
  1em plus 0.5em minus 0.4em\relax Now Publishers Inc, 2009.

\bibitem{StoGeo1-Theory}
F.~Baccelli and B.~Blaszczyszyn, \emph{Stochastic Geometry and Wireless
  Networks, Volume 1: Theory}.\hskip 1em plus 0.5em minus 0.4em\relax NOW, Dec.
  2009.

\end{thebibliography}

\end{document}